\documentclass[conference]{IEEEtran}
\IEEEoverridecommandlockouts
\usepackage{cite}
\usepackage{amsmath,amssymb,amsfonts}
\usepackage{algorithmic}
\usepackage{graphicx}
\usepackage{textcomp}
\usepackage{xcolor}
\def\BibTeX{{\rm B\kern-.05em{\sc i\kern-.025em b}\kern-.08em
    T\kern-.1667em\lower.7ex\hbox{E}\kern-.125emX}}

\usepackage{makecell}
\usepackage{multirow}

\begin{document}

\title{Heart Rate Estimation from Ballistocardiography Based on Hilbert Transform and Phase Vocoder 
}
\author{
\IEEEauthorblockN{Qingsong Xie, Guoxing Wang, Senior Member, IEEE  and Yong Lian, Fellow, IEEE }
\IEEEauthorblockA{Department of Micro-Nano Electronics, Shanghai Jiao Tong University\\
Shanghai, China}
}

\maketitle

\begin{abstract}
This paper presents  a robust method to monitor heart rate (HR)  from BCG (Ballistocardiography) signal, which is acquired from the sensor embedded in a chair or a mattress. The proposed algorithm addresses the shortfalls in traditional Fast Fourier Transform (FFT) based approaches by introducing Hilbert Transform to extract the pulse envelope that models the repetition of J-peaks in BCG signal. The frequency resolution is further enhanced by applying FFT and phase vocoder to the pulse envelope. The performance of the proposed algorithm is verified by experiment from 7 subjects. For HR estimation, mean absolute error (MAE) of 0.90 beats per minute (BPM) and standard deviation of absolute error (STD) of 1.14 BPM are obtained. Pearson correlation coefficient between estimated HR and ground truth HR of 0.98 is also achieved.
\end{abstract}

\section{Introduction}
Cardiovascular diseases (CVDs) are the leading factors of  death according to a report from American Heart Association\cite{benjamin2017heart}. One way to alleviate negative consequence of CVDs is to provide in-home  heart rate (HR) monitoring system that is convenient to use.
Nowadays, various devices for heart rate measurement are available. However, the existing wearable solutions require the use to attach sensors on the body, which may irritate the skin and  usually present discomfort to users\cite{Bandodkar2014Non,Galli2017Robust}. Our interest is to provide a non-intrusive and unobtrusive way to monitor HR in an in-home environment. Ballistocardiography (BCG) signal records the motions of human body generated by sudden ejection of blood into the vessels at every cycle. Such signal contains rich information, which can offer respiration rate and heart rate. An obvious attraction to users of BCG signal is that it can be directly obtained by sensor positioned under chair without any electrode attached to body. It provides a convenient, low cost, easy to use solution for HR monitoring at home. Therefore, BCG-based system becomes  ideal for heart rate monitoring in natural in-home environment.

Many methods have been developed to calculate HR from BCG signal. One clustering approach (CA) was presented to detected J-peak from BCG as heartbeat\cite{Rosales2012Heartbeat}. It extracts three features from BCG signal, including J-peak and adjacent valley, as the input of k-means clustering with the number cluster of 2. The heartbeat class is considered as the cluster with the smaller number of objects while the other one is regarded as non-heartbeat class. Heart rate is calculated by time span between heartbeats. However, it results in erroneous heart rate estimation due to over-detected or missed heartbeats. One short-time energy (SEN) was proposed to estimate heart rate\cite{Lydon2015Robust,Arias2016Unobtrusive}.  The energy of BCG is calculated using short-time energy function. The peaks from energy waveform are located as heartbeats to obtain  heart rate estimates.  Nevertheless, it can not extract good quality of energy waveform when BCG signal shows irregular pattern.  Paalasmaa et al.\cite{Paalasmaa2015Adaptive} proposed an adaptive method for HR estimation. It extractes heartbeat template from BCG and detects the points that best match the template as heartbeats. Heart rate is calculated through consecutive heartbeat intervals in one 30-s window. It needs many parameters to be tuned based on the dataset on which the algorithm is designed and tested. As a result, the performance can not be guaranteed on unknown signal. In \cite{gomez2016novel}, one zero phase bandpass filter with cutoff frequencies of 2 Hz and 7 Hz is first employed to preserve BCG power. BCG envelope is then extracted by applying low-pass Butterworth filter with cutoff frequency of 2 Hz. The peaks of envelope is detected for HR computaion. The envelope  usally appears irregularly when BCG signal dose not have consistent pattern. Pino et al.\cite{Pino2017BCG} put forward multiple smoothed length transform (SLT) method for heart rate estimation from BCG. It selects four fixed windows for SLT computation. This information provides a search window for peak identification. Heart rate is obtained through detected peaks. However, it selects window sizes experimentally, hard to tackle the case with large range of heart rate. All methods above are based on time-domain to calculate heart rate, which are subject to noises caused by body movement.
  
In our work, we present a robust method, Hilbert Transform with Phase Vocoder (HTPV), for HR estimation from BCG sensors. Unlike previous works based on time domain to estimate HR, frequency information is used to estimate HR. We utilize Hilbert Transform (HT) to extract the envelope of BCG signal. Fast Fourier Transform (FFT) is used to calculate frequency spectrum of envelope. Frequency resolution is further increased using phase vocoder (PV) to obtain accurate HR value.

The rest of this paper is organised as follows. Section \uppercase\expandafter{\romannumeral2} describes the proposed method for HR estimation, including Hilbert Transform and phase vocoder. Section \uppercase\expandafter{\romannumeral3} details the experimental setup, metrics for evaluating the accuracy of methods and results.
Conclusion is made in Section \uppercase\expandafter{\romannumeral4}.

\section{Methodology}
\subsection{Overall Architecture}
\begin{figure}[tb]
\centering
\includegraphics[width=0.4\textwidth]{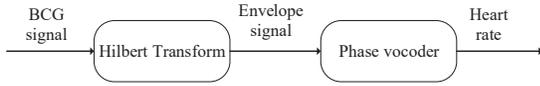}
\caption{The overall architecture of the proposed method }
\label{fig:flow}
\end{figure}

A 30-s sliding window with increment step of 15 s is sliding on BCG signal. HR is estimated in each time window. The proposed method consists of two stages, Hilbert Transform and phase vocoder. Hilbert Transform is used to extract the envelope modelling the repetition of J-peak in BCG signal. The use of phase vocoder is to reduce frequency spacing. Heart rate is finally estimated in frequency domain. The overall architecture is shown in Fig.\ref{fig:flow}.   
 
\subsection{Hilbert Transform}
It is not appropriate  to estimate heart rate from direct Fourier Transform of BCG signal \cite{alametsa2013local}, especially for the BCG recorded from the sensors embedded in chair. This is because the embedded sensor does not generate the significant fundamental frequency corresponding to heart rate. In order to overcome this problem, we propose to use Hilbert Transform. 

The chair-based BCG data is modelled as (\ref{eq:model}), composed of three components.
\begin{equation}
s(n)=r(n)+a(n)cos(2\pi f_0t)+e(n)\label{eq:model}
\end{equation}
$ r(n) $ represents respiration signal described by sinusoid signal with frequency equal to respiration signal. $ a(n)$ is the pulse envelope that models  the repetition of J-peak in BCG signal in the form of periodic pulse sequence. Its frequency spectrum  mainly consists of  harmonics, fundamental frequency of which  corresponds to heart rate. $ a(n) $ can not be observed, only modulated $ a(n) $ with modulation frequency ($ f_0 $) is visible. $ e(n) $ denotes additive noises.

Respiration signal and additive noises can be removed by band-pass filter. In our work, we apply one band-pass filter with cutoff frequencies of 0.7 and 10 Hz since effective frequencies in BCG primarily lie in this range. The filter is constructed in forward and reverse directions to avoid phase distortion. Through this procedure, we get respiration-removed signal, as in (\ref{eq:model1}). 
\begin{equation}
s'(n)=a(n)cos(2\pi f_0t)+e'(n)\label{eq:model1}
\end{equation}

The next step is to extract $ a(n) $ from modulated signal $ a(n)cos(2\pi f_0n) $. In order to achieve this target, Hilbert Transform is used. As we know, HT yields a signal ($ s_j(n) $) that reserves amplitude information and introduces 90 degree phase shift of input signal. Therefore, analytical signal ($ h(n) $) is produced with input signal as real component and output of HT as imaginary component, which is given by (\ref{eq:model2}).

\begin{equation}
h(n)=s'(n)+js_j(n)\label{eq:model2}
\end{equation}
When  the noise is negligible and the frequency of $ a(n) $ is narrow   in contrast to modulation frequency $ f_0 $, $ s_j(n) $ is approximately equal to (\ref{eq:HT}).
\begin{equation}
s_j(n)\approx a(n)sin(2\pi f_0n)\label{eq:HT}
\end{equation}

$ a(n) $ is thus acquired by taking the amplitude of $ h(n) $, given by (\ref{eq:model3}).
\begin{equation}
p(n)=\vert h(n)\vert ^2 =s'(n)^2+s_j(n)^2=a(n)^2\label{eq:model3}
\end{equation}

\begin{figure}[tb]
   \centering
        \includegraphics[height=0.5\textwidth,width=0.5\textwidth]{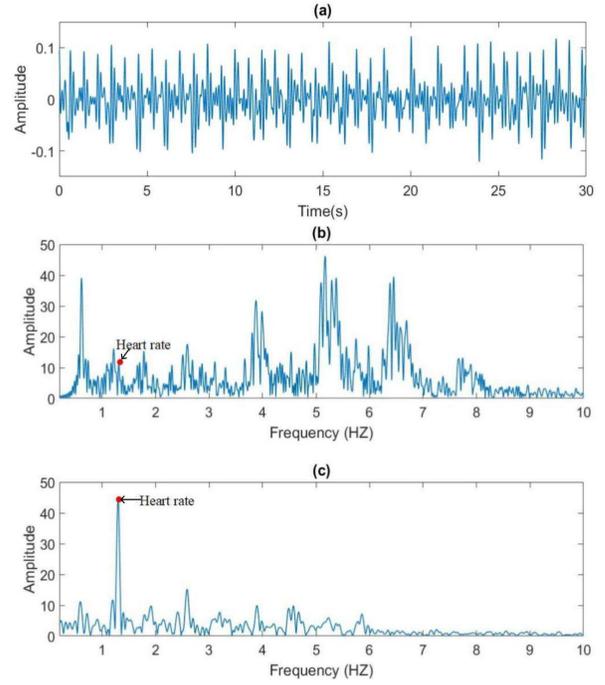}  
    \caption{An example of HT. (a) BCG signal.  (b) Frequency spectrum for BCG. (c) Frequency spectrum for envelope extracted by HT.  }
     \label{Fig:HT}
\end{figure}

Fig.\ref{Fig:HT} gives a representative example of  the improvement of estimating frequency spectrum for HR estimation  after extracting envelope using the proposed HT method. Fig.\ref{Fig:HT} (a) shows a 30-s BCG signal. The frequency spectrum calculated by FFT for BCG is shown in Fig.\ref{Fig:HT} (b). Obviously, the frequency with the largest amplitude deviates quite far from true HR. Therefore, frequency spectrum of BCG signal can not be straightly used to estimate HR.  Fig.\ref{Fig:HT} (c) demonstrates that frequency spectrum of envelope after HT  primarily consists of the frequency corresponding to true HR and its harmonics, which offers effective information for HR estimation.

\subsection{Phase Vocoder}
Fast Fourier Transform is first used to calculate the initial frequency spectrum ($ f_i $) of envelope signal $ a(n) $. However, the frequency resolution of FFT is limited by window size (30 s), which  is equal to $ 1/30*60 =2 $ beat per minute (BPM). Zero-padding before FFT  is thus  used to interpolate spectrum signal to other frequencies so as to decrease frequency spacing between consecutive FFT bins. Apart from zero-padding, phase vocoder\cite{flanagan1966phase,puckette1998accuracy,cotter2005sparse} is applied to further advance frequency resolution. 

Phase vocoder estimates instantaneous frequency based on changes of phase. Phase changes of two successive frames encode the deviation of the true frequency from the bin center frequency. Therefore, instantaneous frequency can be computed using (\ref{eq:PV1}).
\begin{equation}
f(t)=\dfrac{1}{2\pi}\dfrac{d\phi(t)}{dt}\label{eq:PV1}
\end{equation}
 
For discrete time signal, instantaneous frequency can be estimated using phase differencing operation defined in (\ref{eq:PV2}),
\begin{equation}
\hat{f}(n)=\dfrac{1}{2\pi}\dfrac{\phi_2-\phi_1+2\pi n}{t_2-t_1}\label{eq:PV2}
\end{equation}

where $ t_2,t_1 $ are the time stamps of
the current and previous frames, respectively. Here $t_2-t_1 =15 s $, is a window shift.  $ \phi_2,\phi_1 $ are the two phases from two consecutive frames. $ n $ is a positive integer. The series, $ \hat{f}(n) $, is calculated for several $ n $, and the value of $ \hat{f}(n) $ closest to the initial frequency $ f_i $ is selected as refined frequency $ (f_r) $ as in (\ref{eq:PV3}).

\begin{equation}
f_r=\hat{f}(k), \text{where}  \ k=\arg\min_n(\hat{f}(n)-f_i)\label{eq:PV3}
\end{equation}

Heart rate is estimated frame by frame. HR in the first frame is obtained by finding the frequency bin with the largest amplitude  from initial frequency with the largest amplitude. Search range is set as 0.6-4 Hz since it is sufficient for heart rate range. Next heart rate is determined in the range that doest not differ from previous HR estimation more than 10 BPM. This controls HR variations between neighboring frames to be less drastic.  The frequency bin index is first detected corresponding to the largest amplitude. At the same index, refined frequency $ f_r $ is obtained and heart rate is estimated as $ 60*f_r $.

\begin{figure}[tb]
\centering
\includegraphics[width=0.5\textwidth]{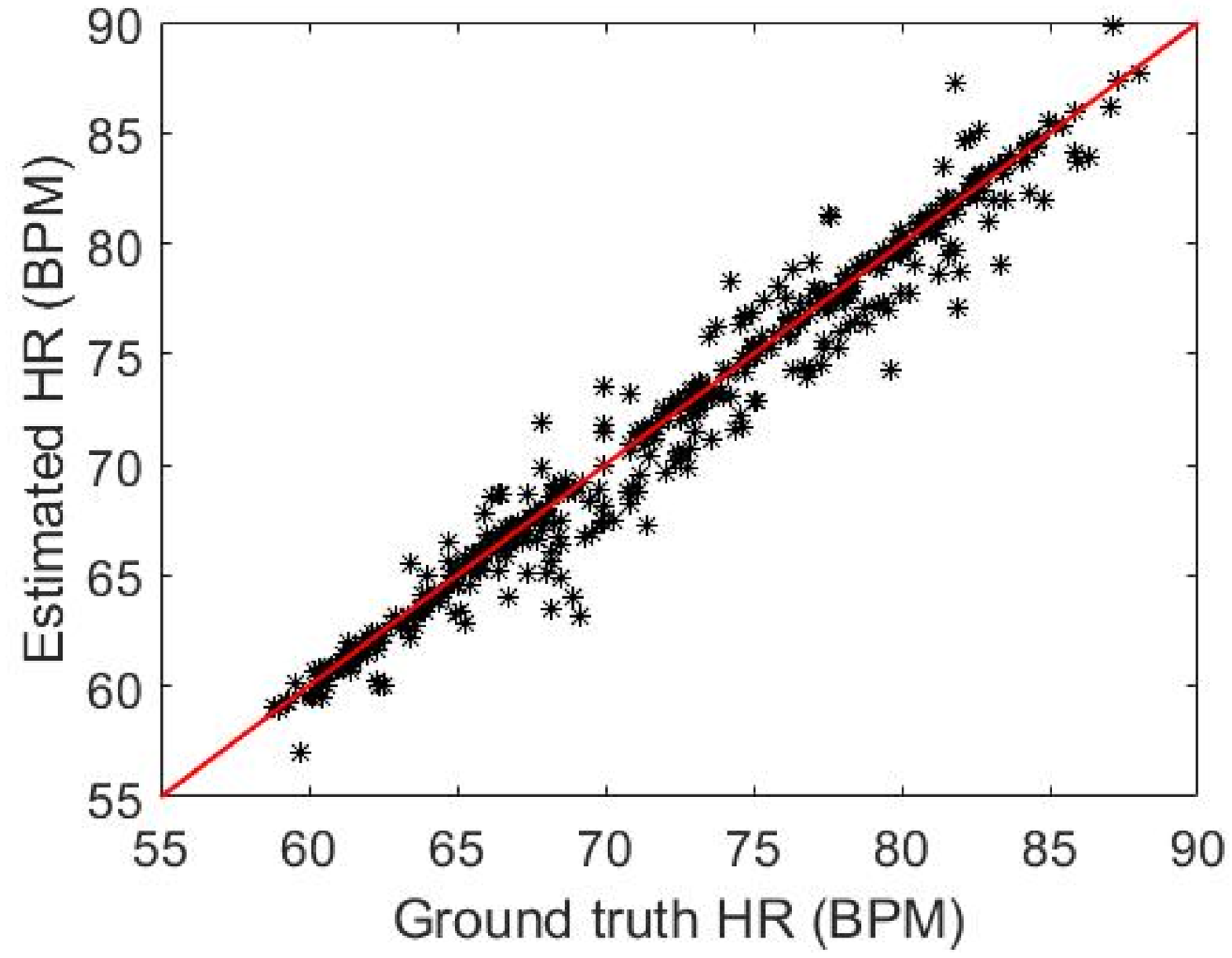}
\caption{Estimated HR versus ground truth HR}
\label{fig:HR}
\end{figure} 

\begin{figure}[tb]
\centering
\includegraphics[width=0.5\textwidth]{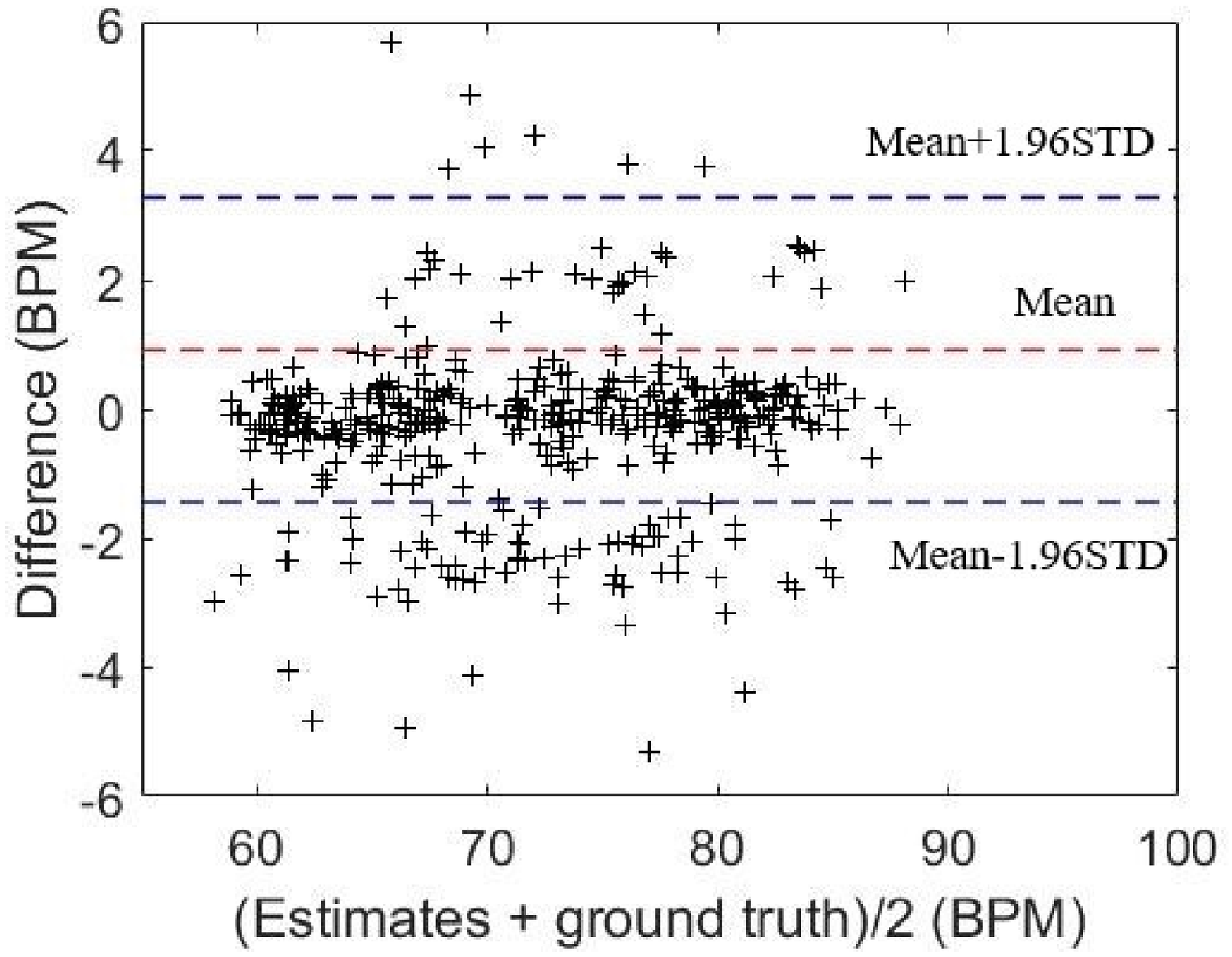}
\caption{Estimated HR versus ground truth HR}
\label{fig:bland}
\end{figure} 
 \begin{table*}[htbp]
 \centering
 \renewcommand\arraystretch{2.0}
 \caption{MAE and STD for heart rate estimation using different methods}
 \begin{tabular}{|p{1.0cm}<{\centering}|p{1.1cm}<{\centering}|p{1.4cm}<{\centering}|p{1.3cm}<{\centering}|p{1.3cm}<{\centering}|p{1.3cm}<{\centering}|p{1.3cm}<{\centering}|p{1.3cm}<{\centering}|p{1.3cm}<{\centering}|p{1.3cm}<{\centering}|}
 \hline
\multirow{2}*{Subject} &\multirow{2}*{Time(Min)}& \multicolumn{2}{c|}{HTPV(proposed)}&\multicolumn{2}{c|}{SEN\cite{Lydon2015Robust}}&\multicolumn{2}{c|}{\cite{gomez2016novel}}&\multicolumn{2}{c|}{SLT\cite{Pino2017BCG}}
 \\
 \cline{3-10}
 ~&~&MAE (BPM)&STD (BPM)&MAE (BPM)&STD (BPM)&MAE (BPM)&STD (BPM)&MAE (BPM)&STD (BPM)
\\
 \hline
 1&18.64&0.37&0.51&0.49&0.58&0.47&0.49&1.12&1.39
 \\
 \hline
 2&13.74&0.91&1.21&1.62&3.82&1.11&0.84&0.78&1.33
\\
\hline
3&14.72&1.07&1.43&1.50&2.10&2.30&2.94&1.87&1.98
\\
\hline
4&18.64&1.16&1.27&2.57&2.82&4.86&7.92&3.83&3.54
\\
\hline
5&14.72&0.67&1.21&2.18&3.73&5.12&8.00&1.50&2.08
\\
\hline
6&17.06&1.07&1.09&1.09&1.70&0.87&1.12&0.82&1.10
\\
\hline
7&13.74&1.06&1.25&1.56&2.33&3.56&4.41&1.32&2.44
\\
\hline
Mean&15.89&0.90&1.14&1.57&2.44&2.61&3.67&1.57&1.90
\\
\hline
 \end{tabular}
\label{table:HR}
\end{table*}
\section{Experimental Results}
\subsection{Experiment Setup} 
We simultaneously recorded one-channel BCG signal and an ECG signal from seven subjects. For every subject, BCG was recorded from chair-based sensor which is located under chair. ECG was also recorded from chest-based electrodes attached to the chest of subject. All signals were sampled with sampling frequency set to 225 Hz. 

During data recording phase, the subject just sat on chair and maintained seating position unchanged. The postures can be changed, for instance, leaning against chair or bowing. The recording lasted approximately 15 minutes for every subject.

\subsection{Performance Metrics}
For each time window, the heart rate from ECG signal is calculated as ground truth. The R-peaks  in ECG are  detected using Pan-Tompkins algorithm\cite{Pan1985A}. After that, the peaks are checked manually to guarantee all R-peaks are accurately detected. Given one frame, we calculate mean R-R time  interval $ T $ and then HR is computed as  $ 60/T $.

Mean absolute  error (MAE) and standard deviation of absolute error (STD) are used to assess the performance of the proposed method, which are given by (\ref{eq:MAD}) and (\ref{eq:STD}).
\begin{equation}
\text{MAE}=\dfrac{1}{N}\sum_{k=1}^{N}\vert BPM_{true}(k)-BPM_{est}(k)\vert\label{eq:MAD}
\end{equation}
\begin{equation}
\text{STD}=\sqrt{\dfrac{1}{N}\sum_{k=1}^N(\vert BPM_{est}(k)-BPM_{true}(k)\vert -\text{MAE)}^2}\label{eq:STD}
\end{equation}
Denoted by $ BPM_{true}(k) $ is the heart rate derived from ECG signal in the $ k $-th frame, and denoted by  $ BPM_{est}(k) $ is the estimated heart rate from BCG signal.

Bland-Altman plot\cite{Bland2007Agreement}, as the third assessment index, is used to measure the agreement between estimates and  ground truth. It gives the difference for every estimate against the average of ground truth and estimate. In our study,  we calculate the mean absolute error $ \pm $ 1.96 standard
deviation of the absolute error ([Mean-1.96STD, Mean+1.96STD]), which contains 95$ \% $ of difference. Pearson correlation coefficient $ r $ is also used to assess accuracy.

\subsection{Results}
Table \ref{table:HR} gives MAE  and STD for every subject using the proposed method.  The methods from \cite{Lydon2015Robust,gomez2016novel,Pino2017BCG}  are implemented as well. MAE and STD using these methods for HR estimation are also listed in Table \ref{table:HR}. It shows that the proposed  HTPV obtains the best accuracy, MAE of 0.90 BPM and STD of 1.14 BPM. Shown in Fig.\ref{fig:HR} is estimated HR from BCG versus ground truth HR from ECG, and  $ r $ of 0.98 between estimated HR and ground truth is achieved. Fig.\ref{fig:bland} shows Bland-Altman plot between the difference and average of ground truth and estimates.

\section{Conclusion}
In this paper, we present a simple but effective method for heart rate estimation based on BCG signal obtained from sensors embedded in a chair.  Hilbert Transform is used to extract pulse envelope. After that, phase vocoder is applied to increase frequency resolution estimated by FFT. Heart rate is finally calculated based on frequency domain. The proposed method achieves MAE of 0.90 BPM, STD of 1.14 BPM, $ r $ of 0.98.  Compared to \cite{Lydon2015Robust,gomez2016novel,Pino2017BCG}, the proposed method has better accuracy.
\bibliographystyle{ieeetr}

\end{document}